\documentclass[12pt]{article}
\usepackage{graphicx}
\usepackage{bm}        
\usepackage{amssymb}   
\usepackage{hyphenat}

\input epsf
\def\beq{\begin{eqnarray}}
\def\eeq{\end{eqnarray}}

\def\ba{\begin{eqnarray}}
\def\ea{\end{eqnarray}}
\def\beq{\begin{eqnarray}}
\def\eeq{\end{eqnarray}}

\def\mpl{M_{\rm Pl}}

\def\E{\mathcal{E}}

\def\p{{\cal P}}

\def\L*{{\cal L}_*}
\def\L{\mathcal{L}}
\def\({\left(}
\def\){\right)}

\def\nn{\nonumber}
\def\p{\partial}
\def\mn{_{\mu \nu}}
\def\stu{St\"uckelberg }
\def\p{\partial}

\def\<{\langle}
\def\>{\rangle}

\def\lsim{\mathrel{\rlap{\lower3pt\hbox{\hskip0pt$\sim$}}
     \raise1pt\hbox{$<$}}}         
\def\gsim{\mathrel{\rlap{\lower4pt\hbox{\hskip1pt$\sim$}}
     \raise1pt\hbox{$>$}}}         

\def\lsim{\mathrel{\rlap{\lower3pt\hbox{\hskip0pt$\sim$}}
     \raise1pt\hbox{$<$}}}         
\def\gsim{\mathrel{\rlap{\lower4pt\hbox{\hskip1pt$\sim$}}
     \raise1pt\hbox{$>$}}}         

\usepackage{amsmath}
\usepackage{amsfonts}
\usepackage{verbatim}
\begin{document}

\begin{titlepage}

\begin{flushright}
{NYU-TH-02/18/11}

\today
\end{flushright}
\vskip 0.9cm

\centerline{\Large \bf Nonlinear Dynamics of 3D Massive Gravity}
\vskip 0.7cm
\centerline{\large Claudia de Rham$^a$, Gregory Gabadadze$^b$, David Pirtskhalava$^b$,}
\centerline{\large Andrew J. Tolley$^c$\ and Itay Yavin$^b$}
\vskip 0.3cm

\centerline{\em $^a$D\'epartment de Physique  Th\'eorique, Universit\'e
de  Gen\`eve,}
\centerline{\em 24 Quai E. Ansermet, CH-1211  Gen\`eve}

\centerline{\em $^b$Center for Cosmology and Particle Physics,
Department of Physics,}
\centerline{\em New York University, New York,
NY, 10003, USA}
\centerline{\em $^c$ Department of Physics, Case Western Reserve University,}
\centerline{\em 10900 Euclid Ave, Cleveland, OH 44106, USA }

\centerline{}
\centerline{}

\vskip 1.cm

\begin{abstract}

We explore the nonlinear classical dynamics of the three-dimensional theory of
``New Massive Gravity" proposed by Bergshoeff, Hohm and Townsend. 
We find that the theory passes  remarkably  highly nontrivial  consistency  checks 
at the nonlinear level. In particular, we  show that:  (1) In the decoupling limit  
of the theory, the interactions of the helicity-0 mode are described by a single cubic
term -- the so-called cubic Galileon --  previously found in the context of the
DGP model and in  certain 4D massive gravities.  (2) The conformal mode of the
metric coincides with the helicity-0 mode in the decoupling limit. Away from this limit
the nonlinear dynamics of the former is described by a certain generalization of Galileon interactions,
which like the Galileons themselves have a well-posed Cauchy problem.  (3)  We give a 
non-perturbative argument based on the presence of additional symmetries that the full 
theory does not lead to any extra degrees of freedom,  suggesting  that
a 3D analog of the 4D  Boulware-Deser ghost is \textit{not} present in this theory.  
Last but not least, we generalize ``New Massive Gravity" and construct  a class of  
3D cubic order massive models that retain the above properties.

\end{abstract}


\end{titlepage}

\newpage

\section{Introduction and Summary}

In this work we study massive gravity in 2+1 spacetime dimensions. 
A well-known theory of this is Topological Massive Gravity  \cite {tmg}.
Here, we focus on the nonlinear dynamics of ``New Massive Gravity"  (NMG)
proposed by Bergshoeff, Hohm, and Townsend, \cite{nmg}. Similar studies on the nonlinear 
structure of massive gravity in {\it 3+1}  dimensions have by now reached a certain level of maturity.
Since these developments are relevant for the motivation and
content  of the present paper, we begin with a brief outline of the most important insights gained.

Fierz and Pauli (FP) \cite{fp} constructed a ghost-free  and  tachyon-less
linear theory of massive spin-2 in Minkowski space, describing 5 degrees of freedom in 4D
(2 degrees of freedom in 3D),  that is consistent with the corresponding
representation  of the Poincar\'e group.  The FP theory has no continuous limit  
to the massless theory \cite{vDVZ}, while the  continuity can be restored in its
nonlinear extensions \cite{Vainshtein}.  However, a generic nonlinear extension  in 4D suffers
from the so-called  Boulware - Deser (BD) ghost:  the Hamiltonian constraint,  that would
restrict the number of degrees of freedom to no more than 5 for any background, is
lost  at the nonlinear level \cite{bd}. The consequences of this  can be transparently
seen in the decoupling limit \cite{AGS}; here,  the problem emerges
in the equation of motion for the helicity-0 mode that exhibits
ill-posed  Cauchy problem   due to  nonlinear terms with
more than  two time derivatives acting on a single field  \cite{Creminelli,Deffayet}.
A toy example of this in 4D is the Lagrangian \cite {AGS}
\beq
\mathcal{L}=\frac{1}{2}\pi\Box\pi+\frac{ (\Box\pi)^3  }{\Lambda^5_5} +\frac{1}{M_P}\pi T,
\label{pilagr}
\eeq
where, $\Lambda_5$ is a certain scale (composed of  the graviton mass and Planck
mass), and $T$ denotes the trace of the external energy-momentum tensor.
The equations of motion that follow from (\ref{pilagr}) involve nonlinear
terms with  more than two time derivatives acting on $\pi$. This leads to an additional (sixth) mode
with negative kinetic term propagating on  nontrivial backgrounds for the helicity-0 mode (e.g., 
that of a collapsing spherical source) \cite{GGruzinov,Creminelli,Deffayet};
this mode is identified with the BD ghost. This follows directly from the standard Ostrogradski 
instability arguments and can be made manifest here by noting that the above theory can be equivalently 
represented by the following lagrangian \cite{Deffayet},
\beq
\mathcal{L'}=\frac{1}{2}\phi \Box\phi -\frac{1}{2}\psi\Box\psi- \epsilon \frac{2}{3\sqrt{3}} 
\psi^{3/2}\Lambda^{5/2}_5+\frac{1}{M_P}\phi
T-\frac{1}{M_P}\psi T.
\label{psilagr}
\eeq
The latter lagrangian is obtained from (\ref{pilagr}) in two steps. 
First one introduces an auxiliary field $\lambda$ and a corresponding lagrangian
$\mathcal{L'}$, which involves no more than single derivatives per field. 
The Lagrangian $\mathcal{L'}$ is defined so that integrating out $\lambda$, one recovers the original
$\pi$-lagrangian (\ref{pilagr}). On the other hand, making the further field 
redefinitions $\lambda^2= \psi$ and $\pi=\phi-\psi$ brings $\mathcal{L'}$ to the form, 
given in (\ref{psilagr}). Here $\epsilon=\pm 1$, depending on the sign of $\Box(\phi-\psi)$.

It is manifest from (\ref{psilagr}) that the additional
degree of freedom $\psi$ is inevitably a ghost; however, its mass is  set by $\langle \psi \rangle ^{-1/2}$, and
although this mass is infinite when expanding around a Minkowski background, which explains why its presence 
was not seen in the original FP analysis, it can drop down to physically accessible scales when expanding around a 
non-trivial background, \cite{Creminelli}.

The simplest example of a nonlinear theory for a helicity-0 mode graviton that does not have
the above problem  was found in the  context of the DGP model
\cite{dgp} in Ref. \cite{lpr}. Its Lagrangian reads as follows:
\beq
\mathcal{L}=\frac{1}{2} \pi\Box\pi+\frac{ (\p\pi)^2\Box\pi  }{\Lambda_3^3}+\frac{1}{M_P}\pi T.
\label{dgp}
\eeq
The specific structure of the cubic term prevents the
appearance of more than two time derivatives in the equations of motion for $\pi$,
thereby rendering it free of the second, ghost-like, degree of freedom on any weak asymptotically flat 
background, \cite{lpr}. The Lagrangian (\ref{dgp}) has been generalized to incorporate the quartic and quintic
self interactions with similar properties  -- the so-called Galileon terms, \cite{nrt}.
These terms have been shown to naturally arise
in probe brane setups \cite{drt} and their extensions  to multi-Galileon
theories have been studied in \cite{DeffayetGalileon,MultiGalileon1,MultiGalileon2}.

More importantly for the present work, Refs.~\cite{drg1,drg2}
showed that a certain class of theories of
4D massive gravity, contain all the Galileon terms
in the decoupling limit.  In these theories, an infinite
number of nonlinear terms can be resummed and a Lagrangian containing just a few terms
can be obtained \cite{drgt}.  Hence, the BD problem can be addressed in  the full-fledged Hamiltonian
approach.  So far the Hamiltonian  has been  calculated  up to and including  quartic order
in nonlinearities and beyond the decoupling limit,
showing  that  the BD problem  does not  arise at this level~\cite{drgt}\footnote{
A potentially dangerous  term  found in the quartic order of the Hamiltonian 
construction in Ref. \cite {Creminelli} has a very special structure \cite {drgt}, 
and due to this,  can be removed to higher orders by a consistent nonlinear field redefinition, 
as shown explicitly in  \cite {drgt}.  The terms that could not have been 
removed, do cancel out automatically in the quartic order 
\cite {drgt}.  We also note that the term of concern of Ref. \cite {Mukhanov},  found in that 
work in  the Lagrangian formalism, is nothing but the term already found in \cite {Creminelli} in the 
Hamiltonian formalism,  that was addressed  above.}.

The purpose of the present work is to both study the nonlinear interactions
of the  helicity-0 mode in the decoupling limit of NMG~\cite{nmg}, determining whether or
not it gives rise to the problematic nonlinear terms, as well as to address the
issue of the BD  ghost in the full theory, away from the decoupling limit.
Since NMG is a 3D theory, one would expect to be able to do more than
in 4D, as well as gain some additional intuition about the 4D constructions.
The divergences arising in 3D being less severe, NMG  can give some valuable insight on the renormalization of gravity, \cite{Oda:2009ys}. Furthermore, a DBI extension was proposed in Ref.~\cite{Gullu:2010pc}, and a remarkable connection with AdS/CFT was established, \cite{Skenderis:2009nt}. The existence of AdS$_3$ Black Hole solutions also makes this theory especially interesting, \cite{Clement:2009gq}.

In this work, we will focus instead on the stability of the theory and show that 
the interaction Lagrangian for the helicity-0 in the decoupling limit of NMG
reduces to  the three-dimensional version of Lagrangian ~(\ref {dgp}), supplemented by an 
equation which determines the tensor modes.  Hence, the theory
has a well posed Cauchy problem. The nonlinear term in (\ref {dgp}) seems to arise universally: It appears
in the context of the 5D DGP  model, as found in \cite{lpr}, in 4D massive gravity \cite{drg1,drg2,drgt}, and now in 3D NMG.
In what follows we demonstrate that in NMG, the helicity-0 mode coincides with the conformal mode of the metric
in the decoupling limit.  This motivates us to study the dynamics of the conformal mode
in the full theory. In ordinary GR, it is well known that the conformal mode has the wrong-sign kinetic term. 
This is not a problem in GR since this mode is not dynamical (it can be removed via the gauge freedom and constraints). 
In NMG the conformal mode is dynamical, but the
GR term itself has a wrong sign\footnote{This is acceptable as there are no propagating
degrees of freedom  in 3D massless GR.}, hence, the conformal mode obtains a correct sign kinetic term.
Furthermore, the nonlinear terms of NMG give rise to interactions for the conformal mode that
are straightforward generalizations of the Galileon interactions, implying that they have a well-defined Cauchy problem.
Moreover, we give a non-perturbative counting of degrees of freedom that demonstrates that in NMG, even away from the 
decoupling limit or from conformally
flat metric configurations, there are no extra modes beyond the two of a massive 3D graviton. This confirms that the 
3D analogue of the BD ghost does not arise nonlinearly.
Last but not least, we present a class of 3D cubic order theories that generalize the NMG construction.

The paper is organized as follows. Section 2 serves as a review where we present some
well-known properties of massive gravity in 4D, adapted to three spacetime dimensions.
We illustrate that all known potential problems that arise when attempting to give a 4D graviton mass,
persist in generic three dimensional models as well. The reader who is well acquainted with the technology and 
issues involved is encouraged to skip directly to section 3 where we investigate the nonlinear dynamics of NMG. 
We start by studying the decoupling limit of the theory, and arrive at a well-defined,
ghost-free theory. We then move to the dynamics of the conformal mode, and find that it coincides
with the helicity-0 graviton in the decoupling limit. We show next that  for conformally flat  metrics the full 
theory is free of any ghosts.  Moreover, we give the non-perturbative arguments in favor of no-ghost propagation 
beyond any limit or approximation. In section 4 we look at NMG from
the perspective of a generalization of the linear FP  model, demonstrating how exactly the
cancellation of the BD ghost happens at the cubic level. Finally, we construct a class of generalizations
of the cubic theory, which exhibit similar properties.

\section{Ghosts and Strong Coupling in Massive Gravity}

In this section we review some of the known results from massive gravity in 4D and introduce the formalism used in this paper. 
We begin by analyzing the Fierz-Pauli model at the linear level. We then extend the discussion to include nonlinear terms.

\subsection{Linear analysis}

The FP model is the unique theory at the linearized level, free of ghosts or tachyons  \cite{Nieu} and propagating only the two degrees of freedom of a massive spin-2 particle in 3D. In terms of the metric perturbation $h\mn\equiv g\mn-\eta\mn$, the FP Lagrangian is given by,
\beq
\mathcal{L}=-\frac{1}{2}M_P h^{\mu\nu}\mathcal{E}_{\mu\nu}^{\rho\sigma}h_{\rho\sigma}-\frac{1}{4}M_P m^2(h^{\mu\nu}h_{\mu\nu}-h^2)+\frac{1}{2}h^{\mu\nu}T_{\mu\nu},
\label{fp}
\eeq
where $m$ sets the graviton mass, $T\mn$ denotes an external energy-momentum source, and $\mathcal{E}_{\mu\nu}^{\rho\sigma}$ is the linear Einstein operator
\beq
\mathcal{E}_{\mu\nu}^{\rho\sigma}h_{\rho\sigma}=- {1\over 2} ( \square h_{\mu \nu} - \partial_\mu \partial^\alpha
h_{\alpha\nu} - \partial_\nu \partial^\alpha
h_{\alpha\mu} + \partial_\mu \partial_\nu h -
\eta_{\mu\nu} \square h + \eta_{\mu\nu}
\partial_\alpha  \partial_\beta h^{\alpha\beta}),
\eeq
with all indices contracted with the flat metric.

In order to single out the propagating helicity-1 and helicity-0 modes of the massive graviton, the metric perturbation can be conveniently decomposed at the linear order in the following way,
\beq
h_{\mu\nu}=\frac{\bar h_{\mu\nu}}{\sqrt{M_P}}+\frac{\partial_\mu V_\nu}{\sqrt{M_P}m^2}+\frac{\partial_\nu V_\mu}{\sqrt{M_P}m^2}, \quad V_\mu= m A_\mu+\partial_\mu \pi,
\eeq
where $\bar h_{\mu\nu}$, $A_\mu$ and $\pi$ encode the (canonically normalized) helicity-2, 1 and 0 components, respectively (note that in 3D the helicity-2 mode does not propagate). With this field content, the theory is invariant under the usual linearized diffeomorphisms
\beq \bar h_{\mu\nu}\rightarrow \bar h_{\mu\nu}+\partial_\mu\xi_\nu+\partial_\nu\xi_\mu, \quad  A_\mu\rightarrow A_\mu-m\,\xi_\mu~,
\eeq
as well as an additional abelian U(1) symmetry, under which the vector and scalar modes transform as,
\beq
A_\mu\rightarrow A_\mu+\partial_\mu \zeta, \quad \quad \pi\rightarrow \pi-m\,\zeta. ~~~~~~
\label{gt}
\eeq
The latter invariance, which at this stage is introduced somewhat artificially, will turn out to provide a very convenient bookkeeping tool when studying different decoupling limits of nonlinear massive gravity.

As remarked in the introduction, the linear FP theory does not possess a continuous massless limit. To see this, we note that in terms of the different helicity components, the $m\rightarrow 0$ limit of the theory is given (up to a total derivative) by the following expression,
\beq
\mathcal{L}_{m=0}=-\frac{1}{2}\bar h^{\mu\nu}\mathcal{E}_{\mu\nu}^{\rho\sigma}\bar h_{\rho\sigma}-\frac{1}{4}F^{\mu\nu}F_{\mu\nu}-\bar h^{\mu\nu}(\partial_\mu\partial_\nu \pi-\eta_{\mu\nu}\Box\pi)+\frac{1}{2\sqrt{M_P}}\bar h^{\mu\nu}T_{\mu\nu},
\label{lim1}
\eeq
where $F_{\mu\nu}$ denotes the usual abelian field strength for $A_\mu$. The mixing between the tensor and scalar modes in (\ref{lim1}) can be eliminated by a linear conformal redefinition
\beq
\bar h_{\mu\nu}= \tilde h_{\mu\nu}+2\eta_{\mu\nu}\pi,
\label{cs}
\eeq
which brings (\ref{lim1}) to the following form
\beq
\mathcal{L}_{m=0}=-\frac{1}{2}\tilde h^{\mu\nu}\mathcal{E}_{\mu\nu}^{\rho\sigma}\tilde h_{\rho\sigma}-\frac{1}{4}F^{\mu\nu}F_{\mu\nu}+2 \pi\Box\pi+\frac{1}{2\sqrt{M_P}}\tilde h^{\mu\nu}T_{\mu\nu}+\frac{1}{\sqrt{M_P}}\pi T.
\label{lim2}
\eeq
It is clear from the latter Lagrangian that the helicity-0 part of the massive graviton does not decouple from matter even in the $m\rightarrow 0$ limit, leading to the famous vDVZ discontinuity \cite{vDVZ}. This is an $\mathcal{O}(1)$ modification of the gravitational interactions as compared to GR, at least in the regime of validity of the linear approximation. However, as first pointed out in \cite{Vainshtein}, in nonlinear theories of massive gravity this approximation typically breaks down at a parametrically large distance $r_V$ from localized sources, allowing for their phenomenological viability. In fact, it is the nonlinear dynamics of the scalar mode itself that screens its contribution to the gravitational potential within the Vainshtein radius, $r_V$, restoring agreement with GR. In a generic nonlinear extension of the FP theory, however, the same nonlinear self-interactions of $\pi$ are responsible for a number of theoretical problems, such as the propagation of ghosts and the ill-posedness of the Cauchy problem, strong coupling at parametrically low scales, and etc. Before turning to the discussion of theories that avoid these problems, it is instructive to have a closer look at their origin by considering the minimal nonlinear extension of the FP model, to which we turn next.

\subsection{Fierz-Pauli model at the nonlinear level}

Let us ``naively" continue the theory of the previous subsection at the nonlinear level by keeping the graviton potential unchanged, while completing the derivative self-interactions to the full Einstein-Hilbert action,
\ba
{\cal L} =  M_{\rm P} \sqrt{-g}  R -   \frac{ M_{\rm P} m^2}{4}
 \(h^2_{\mu\nu}-h^2\).
\label{PF}
\ea
Here index contractions on the metric perturbation are performed using the flat metric $\eta^{\mu\nu}$. In order to single out the high-energy degrees of freedom in (\ref{PF}), we employ a trick \textit{\`a la} \stu\cite{Stueckelberg:1957zz}: For the purpose of dealing with the theory at cubic order, it is more convenient to restore general covariance by introducing the decomposition for $h_{\mu\nu}$, patterned after the nonlinear gauge transformation of the metric perturbation
\beq
h_{\mu\nu}=\frac{\bar h_{\mu\nu}}{\sqrt{M_P}}+\frac{\p_\mu V_\nu+\p_\nu V_\mu}{\sqrt{M_P} m^2}+\frac{\p_\mu V^\alpha \p_\nu V_\alpha}{M_P m^4},
\label{nls}
\eeq
where $\bar h_{\mu\nu}$ and $V_\mu$ are further decomposed as
\beq
\bar h_{\mu\nu}=\hat h_{\mu\nu}+\frac{\p_\mu V^\alpha \hat h_{\alpha\nu}+\p_\nu V^\alpha \hat h_{\alpha\mu}+V^\alpha\p_\alpha \hat h_{\mu\nu}}{\sqrt{M_P} m^2}, \quad V_\mu= m A_\mu+\p_\mu \pi.
\label{dec'}
\eeq
The fields $\tilde h_{\mu\nu}=\hat h_{\mu\nu}-2 \eta_{\mu\nu}\pi$, $A_\mu$ and $\pi$ will describe the canonically normalized tensor, vector, and scalar modes at high energies, respectively.
Plugging the decomposition (\ref{dec'}) into the Lagrangian (\ref{PF}), and considering the limit
\beq
M_p\rightarrow \infty ,\quad m\rightarrow 0, \quad (M_P^{1/2} m^4)^{2/9}\equiv \Lambda_{9/2}=\text{finite},
\eeq
one recovers a particular high-energy (``decoupling") limit of the theory. Performing the conformal shift (\ref{cs}) of the helicity-2 mode, (\ref{PF}) reduces up to a total derivative to the following expression,
\beq
\mathcal{L}_{dec}=-\frac{1}{2}\tilde h^{\mu\nu}\mathcal{E}_{\mu\nu}^{\rho\sigma}\tilde h_{\rho\sigma}-\frac{1}{4}F^{\mu\nu}F_{\mu\nu}+2 \pi\Box \pi+\frac{1}{2 \Lambda ^{9/2}_{9/2}}\left((\Box\pi)^3-(\Box\pi)(\partial_\mu\partial_\nu\pi)^2\right)\nonumber \\
+\frac{1}{2\sqrt{M_P}}\tilde h^{\mu\nu} T_{\mu\nu}+\frac{1}{\sqrt{M_P}}\pi T.
\label{dll}
\eeq
 As noted in the introduction, the cubic decoupling limit of the FP theory is amended by the helicity-0 self-interactions of the form $(\partial^2\pi)^3$. They successfully screen the scalar contribution to the gravitational potential of a point source of mass $M$ inside the Vainshtein radius $r_V$ , given by
\beq
r_V=\( \frac{M}{M_Pm^4} \)^{1/4},
\eeq
which restores the agreement with General Relativity. This however happens at the expense of introducing a ghost in the theory: Although infinitely heavy on a flat background, the mass of the ghost becomes low enough to be disruptive around any reasonably localized source \cite{AGS, Creminelli, Deffayet}\footnote{Although the analysis in these references is performed in four spacetime dimensions, we confirmed that the conclusions remain intact in (2+1)D as well.}. Even if one does not consider an external source, the cubic self-interactions of the scalar field, containing more than two time derivatives, cause the Cauchy problem to be ill-posed for the theory at hand.

Recently, a class of theories of (four-dimensional) massive gravity that avoid these problems have been constructed \cite{drg1,drg2,drgt}. These theories modify the graviton potential order-by order, so as to cancel all potentially dangerous self-interactions of the scalar field. As a result, one obtains a sensible effective field theory for a massive graviton, with the cutoff given by $\Lambda_3=(M_P m^2)^{1/3}$ in four spacetime dimensions. Moreover, the decoupling limit of the theory, obtained by keeping $\Lambda_3$ finite while setting the graviton mass and Planck constant to zero and infinity respectively, is free of ghosts below the cutoff. Furthermore, the form of the decoupling limit is \textit{unique} at the quartic order; in other words, any nonlinearities in the potential of order higher than the quartic one have no effect on it\footnote{Interestingly enough, a recently proposed nonlinear completion of Fierz-Pauli massive gravity \cite{g, dr} automatically produces exactly the right structure of the potential at the cubic level, so as to fall into the category of such models, \cite{drg1}.}.

\section{New Massive Gravity}

In this section we analyze the nonlinear dynamics of NMG. As we show at the end of the section, viewed as a generalization of the FP model, NMG represents a curious example. In addition to introducing a potential for the graviton, it also modifies its derivative self-interactions beyond the ones originating from the Einstein-Hilbert action, leading to a well-defined decoupling limit similar to the four-dimensional case of \cite{drg1,drg2,drgt}. Most importantly, however, the full theory can be shown to be free of the BD ghost. We turn to the discussion of this remarkable property next.

One way to define NMG is through the following action
\beq
S_{NMG}=M_P\int \,d^3x \ \sqrt{g} \left[-R-f^{\mu\nu}G_{\mu\nu}-\frac{1}{4}m^2(f^{\mu\nu}f_{\mu\nu}-f^2)\right].
\label{nmg}
\eeq
Here $G_{\mu\nu}$ and $R$ are the usual Einstein tensor and Ricci scalar for the metric $g_{\mu\nu}$, respectively. All the indices are raised with the inverse metric $g^{\mu\nu}$ and $f_{\mu\nu}$ represents an auxiliary covariant tensor field ($f\equiv g^{\mu\nu}f_{\mu\nu}$) that can be integrated out to yield a particular higher-derivative extension of the `wrong sign' GR \cite{nmg}\footnote{Note the minus sign in front of the Einstein-Hilbert term in (\ref{nmg}), which is characteristic of Topologically Massive Gravity as well \cite{tmg}. }. The equations of motion, obtained by varying (\ref{nmg}) with respect to $g_{\mu\nu}$ and $f_{\mu\nu}$ (we keep the overall factors of $\sqrt{g}$ which will be essential for the perturbative analysis of the equations beyond the linear level) are given by the following expressions
\begin{align}
\sqrt{g}G^{\mu\nu}=\frac{1}{2}\sqrt{g}g^{\mu\nu}f^{\alpha\beta}G_{\alpha\beta}-\sqrt{g}f^\mu_\alpha G^{\alpha\nu}-\sqrt{g}f^\nu _\alpha G^{\alpha\mu}+ \frac{1}{2}\sqrt{g}(fR^{\mu\nu}-f^{\mu\nu}R)\nonumber \\+
\frac{1}{2}\sqrt{g}(\nabla^\alpha\nabla^\mu f^\nu_\alpha + \nabla^\alpha\nabla^\nu f^\mu_\alpha-\Box f^{\mu\nu}-g^{\mu\nu}\nabla^\alpha\nabla^\beta f_{\alpha\beta}-\nabla^\mu\nabla^\nu f+g^{\mu\nu}\Box f) \nonumber \\ +\frac{1}{8}m^2\sqrt{g}g^{\mu\nu}(f^{\alpha\beta}f_{\alpha\beta}-f^2)-\frac{1}{2}m^2\sqrt{g}(f^\mu_\alpha f^{\alpha\nu}-f^{\mu\nu}f)\, ,
\label{eq1}
\end{align}
\begin{align}
\sqrt{g}G^{\mu\nu}+\frac{1}{2}\sqrt{g}m^2(f^{\mu\nu}-g^{\mu\nu}f)=0,
\label{eq2}
\end{align}
and admit a Minkowski vacuum with $g_{\mu\nu}=\eta_{\mu\nu}$ and $f_{\mu\nu}=0$ (the Einstein and Ricci tensors are taken as functions of the metric $g_{\mu\nu}$ in the latter equations). From Eq.~(\ref{eq1}) we see that at the linear level the perturbations $h_{\mu\nu}$ and $f_{\mu\nu}$ of the metric and the auxiliary field satisfy
\beq
G^{(1)}_{\mu\nu}(h-f)=0,
\nonumber
\eeq
where $G^{(1)}_{\mu\nu}$ denotes the linearized Einstein tensor. Since it does not propagate any degrees of freedom in three dimensions, the only solution to the latter equation is (up to a gauge, which we choose to be zero) $h_{\mu\nu}=f_{\mu\nu}$. This, in combination with Eq.~(\ref{eq2}), yields the usual FP equations of motion for $f_{\mu\nu}$ in the linearized approximation.

\subsection{Exact decoupling limit of NMG}

Before turning to the full theory, we start by considering the decoupling limit of NMG,  defined as,
$$M_P\to \infty, \quad m\to 0, \quad \Lambda_{5/2}\equiv (\sqrt{M_p}m^2)^{2/5}=\text{fixed}.$$
We consider the (\textit{\`a posteriori} justified) ansatz
\beq
h\mn=\frac{\bar h\mn}{\sqrt{M_P}}, \quad  f\mn=\frac{\bar f\mn}{\sqrt{M_P}}+\nabla_{\mu}V_{\nu}+\nabla_{\nu}V_{\mu},
\eeq
where $V_\mu$ is a vector field which will encode the helicity-0 and -1 degrees of freedom of the massive three-dimensional graviton, and $\nabla$ denotes the covariant derivative associated with the metric $g\mn$. Plugging the latter decomposition into  (\ref{nmg}),  we obtain, up to a total derivative, the following expression for the Lagrangian of NMG,
\ba
\mathcal{L}_{NMG} &=& M_P \sqrt{g} \left[-R-\frac{1}{\sqrt{M_P}}\bar{f}^{\mu\nu} G\mn-\frac{m^2}{4M_P} (\bar{f}^{\mu\nu}\bar{f}\mn-\bar{f}^2) \right. \nn \\&-& \left. \frac{m^2}{\sqrt{M_P}}\bar{f}^{\mu\nu}(\nabla_\mu V_\nu -g\mn\nabla^\alpha V_\alpha)
-\frac{m^2}{2}(\nabla^\mu V^\nu \nabla_\mu V_\nu-\nabla^\mu V_\mu \nabla^\nu V_\nu) \right. \nn \\ &+&\left. \frac{m^2}{2}R^{\mu\nu}V_\mu V_\nu \right].
\label{exactlagr}
\ea
In here, the last term comes from the commutator of covariant derivatives and the quantities $R$ and $G\mn$, as well as the covariant derivative itself, are associated with $g\mn$. Decomposing $V_\mu$ further into the canonically normalized helicity-1 and -0 modes $A_\mu$ and $\pi$,
\beq
V_\mu=\frac{A_\mu}{\sqrt{M_P}m}+\frac{\nabla_\mu\pi}{\sqrt{M_P}m^2},
\eeq
one arrives at the Lagrangian, which up to a total derivative can be written as,
\ba
\mathcal{L}_{NMG}&=&\sqrt{g}\left[-M_P R-\sqrt{M_P} \bar f^{\mu\nu}G\mn -\frac{1}{4}m^2(\bar{f}^{\mu\nu}\bar{f}\mn-\bar{f}^2)-m \bar{f}^{\mu\nu}(\nabla_\mu A_\nu \right. \nn \\ &-& \left. g\mn\nabla^\alpha A_\alpha) -\bar{f}^{\mu\nu}(\nabla_\mu\nabla_\nu\pi-g_{\mu\nu}\Box\pi)-\frac{1}{2}(\nabla^\mu A^\nu\nabla_\mu A_\nu-\nabla^\mu A_\mu \nabla^\nu A_\nu)  \right. \nn \\ &+& \left. \frac{1}{2}R^{\mu\nu}A_\mu A_\nu +\frac{2}{m}R^{\mu\nu}\nabla_\mu\pi A_\nu+\frac{1}{m^2}R^{\mu\nu}\nabla_\mu\pi \nabla_\nu\pi\right]. \nn
\ea
It is straightforward to check that at the linearized level the theory propagates the two degrees of freedom (encoded in $A_\mu$ and $\pi$) of the massive 3D graviton.
At this point we take the exact decoupling limit $M_{P} \rightarrow \infty$
defined by keeping the scale $\Lambda_{5/2}$ fixed, giving rise to the following Lagrangian,
\ba
\mathcal{L}_{NMG}^{5/2}&=&\frac{1}{2}\bar h^{\mu\nu}(\mathcal{E}\bar h)\mn-\bar f^{\mu\nu}(\mathcal{E}\bar h)\mn-\bar f^{\mu\nu}(\p_\mu\p_\nu\pi-\eta\mn\Box\pi)-\frac{1}{4}F^2\mn\nn \\ &+&\frac{1}{\Lambda^{5/2}_{5/2}}\bar R^{(1)\mu\nu}\p_\mu\pi\p_\nu\pi,
\ea
where $\bar R^{(1)}\mn$ denotes the linearized Ricci tensor for $\bar h\mn$, $F\mn$ is the usual abelian field strength for $A_{\mu}$ and all indices are contracted with the flat metric. The auxiliary field $\bar f\mn$, being a Lagrange multiplier in the limit at hand, imposes the constraint
\beq
(\mathcal{E}\bar h)\mn=-\p_{\mu}\p_{\nu}\pi+\eta\mn\Box\pi,
\eeq
which is solved by,
\beq
\bar h\mn=2\pi \,  \eta\mn + \text{gauge transformation}.
\eeq
 Plugging the latter expression back into the Lagrangian and extracting a total derivative, one finally obtains the exact decoupling limit of the theory,
 \beq
 \mathcal{L}_{NMG}^{5/2}=2 \pi\Box\pi-\frac{1}{2}(\p\pi)^2\Box\pi-\frac{1}{4}F^2\mn.
 \label{dlpi}
 \eeq
The decoupling limit of NMG therefore contains the helicity-0 mode with the cubic Galileon self-interaction, together with a free helicity-1 mode! The tensor modes $h_{\mu \nu}$ and $f_{\mu\nu}$ have both disappeared in this limit because they become massless, and by standard arguments massless spin 2 fields in 3D carry no propagating degrees of freedom (although in the presence of sources they will still contribute to the interaction energy).
Since $\bar h\mn=2\pi \,  \eta\mn +$ gauge transformation, we automatically generate a coupling to matter that is of the form $\propto \pi \, T$ if external sources are considered. Remarkably, as we show below, this is an example of a theory of a massive 3D graviton with modified derivative self-interactions, free of ghosts to all orders in the decoupling limit. Even more importantly, we show next that the absence of the BD ghost persists even away from this limit.

\subsection{Dynamics of the conformal mode}

Before presenting arguments in support of this last statement, as a preliminary step it is very instructive to have a look at the dynamics of the conformal mode in NMG. Integrating out the auxiliary field $f\mn$ in (\ref{nmg}), one arrives at the original representation of the NMG action~\cite{nmg},
\beq
\mathcal{L}_{NMG}=M_P \sqrt{g}\left[-R+\frac{1}{m^2}\left( R^{\mu\nu}R\mn-\frac{3}{8}R^2 \right)   \right].
\label{nmg'}
\eeq
Considering conformally flat configurations of the metric
\beq
g\mn=\Omega^2 \eta\mn, \qquad \Omega=e^{\phi/\sqrt{M_P}},
\eeq
the Lagrangian (\ref{nmg'}) can be rewritten in terms of the field $\phi$, as
\ba
\mathcal{L}^{c}_{NMG}&=&e^{\phi/\sqrt{M_P}}\( 4\sqrt{M_P}~\Box\phi+2(\p_\mu\phi)^2 \)+e^{-\phi/\sqrt{M_P}}~ [~\frac{1}{m^2} \( (\p_\mu\p_\nu\phi)^2-(\Box\phi)^2 \)\nn \\ &-&\frac{2}{\sqrt{M_P}m^2}\p^\mu\p^\nu\phi\p_\mu\phi\p_\nu\phi+\frac{1}{2M_P m^2}(\p_\mu\phi)^2(\p_\nu\phi)^2~],
\label{conformalnmg}
\ea
where all indices are contracted with the flat metric. Curiously enough, the $\Lambda_{5/2}$ decoupling limit of this theory coincides in form with the decoupling limit Lagrangian of the  helicity-0 mode in Eq.~(\ref{dlpi})
\beq
\mathcal{L}^{c}_{NMG} \supset 2 \phi\Box\phi-\frac{1}{2}(\p\phi)^2\Box\phi,
\eeq
identifying the helicity-0 part of the massive graviton with the conformal mode in this limit.
One can show that even away from the decoupling limit, the theory given by (\ref{conformalnmg}) does not propagate ghosts as one would naively assume from the presence of higher derivative interactions.

Indeed, up to a total derivative, Eq.~(\ref{conformalnmg}) can be rewritten as,
\ba
\mathcal{L}^{c}_{NMG}&=&e^{\phi/\sqrt{M_P}}\( 4\sqrt{M_P}~\Box\phi+2(\p_\mu\phi)^2 \)+\frac{1}{2m^2} e^{-\phi/\sqrt{M_P}}\( (\p_\mu\p_\nu\phi)^2-(\Box\phi)^2 \)\nn \\ &+& \frac{1}{2\sqrt{M_P}m^2}e^{-\phi/\sqrt{M_P}}\p^\mu\phi\p^\nu\phi\p_\mu\p_\nu\phi.
\label{conformalnmg'}
\ea
Naively, the second and third terms in the above Lagrangian, involving more than two time derivatives, might lead to the presence of ghosts (or equivalently, the ill-posedness of the Cauchy problem) in the theory. One can however show that the specific structure of these operators makes them harmless. The second term is conveniently expressed in terms of the Levi-Civita symbol as follows,
\beq
\propto  e^{-\phi/\sqrt{M_P}} \varepsilon^{\mu\alpha\rho\sigma}\varepsilon_{\nu\beta\rho\sigma}~\p_\mu\p^\nu\phi~\p_\alpha\p^\beta\phi.
\eeq
The antisymmetry of the Levi-Civita symbol can then be used to show that no terms with more than two time derivatives are present in the equations of motion. Similarly, noticing that the third term in Eq.~(\ref{conformalnmg'}) includes a factor of a peculiar form $$\p^\mu\phi\p^\nu\phi\p_\mu\p_\nu\phi,$$ it becomes fairly straightforward to show that it does not cause the Cauchy problem to be ill-posed either. Indeed, this latter expression, being (up to a total derivative) equivalent  to the DGP galileon $\Box\phi(\p_\mu\phi)^2$, is well known to lead to no more than two time derivatives per field in the equation of motion. As can be straightforwardly checked, the factor of $e^{-\phi/\sqrt{M_P}}$ in front does not alter this property.

\subsection{No ghosts in New Massive Gravity}

Finally, we give an argument demonstrating the absence of the BD ghost in the full NMG. The argument we shall give is based on counting non-perturbatively the degrees of freedom utilizing the symmetries. The key observation is that there is a non-perturbative generalization of the $U(1)$ symmetry described in Eq.~(\ref{gt}) which combined with the existing 3D reparameterization invariance is sufficient to demonstrate that only the 2 propagating degrees of freedom remain non-perturbatively.

The Lagrangian (\ref{exactlagr}) is invariant under the usual reparametrizations of coordinates (3D diffeomorphisms), under which the metric $g\mn$ and the auxiliary field $\bar f\mn$ transform as tensors, while $V_\mu$ is a vector. One can use this symmetry to eliminate all potentially propagating degrees of freedom in the 3D metric $g\mn$.
By introducing the vector field $V_{\mu}$ we also introduce a new three-parameter local invariance (which we refer to as secondary linearized 3D diffeomorphisms), under which the metric $g\mn$ remains unchanged, while $\bar f\mn$ and $V_\mu$ transform as,
\beq
\bar f\mn\to \bar f\mn+\sqrt{M_P}(\nabla_\mu \xi_\nu+\nabla_\nu \xi_\mu), \qquad V_\mu\to V_\mu-\xi_\mu.
\label{sym2}
\eeq
It will prove helpful to slightly rewrite the Lagrangian (\ref{exactlagr}) in the following form,
\ba
\mathcal{L}_{NMG}&=&\sqrt{g}~\left[-M_P R-\sqrt{M_P}\tilde f^{\mu\nu}G\mn-\frac{1}{4}m^2(\tilde{f}^{\mu\nu}\tilde{f}\mn-\tilde{f}^2)-\frac{1}{4}\bar F^{\mu\nu} \bar F\mn \right. \nn \\
&-&\left. \frac{1}{4}\frac{m^2}{\sqrt{M_P}}\tilde f^{\mu\nu}(\bar V_\mu \bar V_\nu+g\mn\bar V^\alpha \bar V_\alpha)-m\tilde f^{\mu\nu} (\nabla_\mu \bar V_\nu-g\mn\nabla^\alpha \bar V_{\alpha})\right. \nn \\ &-&\left. \frac{1}{2}\frac{m}{\sqrt{M_P}}\nabla^\mu \bar V^\nu (\bar V_\mu \bar V_\nu+g\mn\bar V^\alpha \bar V_\alpha)+\frac{1}{8}\frac{m^2}{M_P}(\bar V^\alpha \bar V_\alpha)^2~\right],
\label{lagr2}
\ea
where the following set of field redefinitions has been used,
\beq
\bar V_\mu=\sqrt{M_P} m V_\mu, \qquad \tilde f\mn=\bar f\mn-\frac{1}{2\sqrt{M_P}}(V_\mu V_\nu-g\mn V^\alpha V_\alpha),
\eeq
and $\bar F\mn$ denotes the abelian field strength for $\bar V_\mu$. The symmetry (\ref{sym2}) induces the corresponding transformation on $\tilde f\mn$, that leaves the action invariant. One can use this freedom to eliminate all potentially propagating d.o.f's in $\tilde f\mn$, leaving $V_\mu$ the only propagating field in the theory.

Generically, $\bar V_\mu$ would propagate three degrees of freedom in three dimensions; however, as we shall argue below, NMG is special in this sense, propagating only two degrees of freedom in the latter field.
The only place in the Lagrangian (\ref{lagr2}) where the time derivative of $\bar V_0$ appears, is the last term of the second line and the first term on the last line, which however include $\dot {\bar V}_0$ only \textit{linearly}. This means that the canonical momentum, conjugate to $\bar V_0$ is independent of $\dot{\bar V}_0$ itself,  enforcing the primary constraint on the system. This reduces the number of propagating degrees of freedom in $\bar V_\mu$ to two, which, due to the arguments presented above, means that these are the only two dynamical degrees of freedom in the full theory of NMG. To make this argument clearer we can define
\beq
\bar V_{\mu}=  A_{\mu}+ \partial_{\mu} \pi ,
\eeq
where $A_{\mu}$ is a vector under usual 3D diffeomorphisms, and transforms as $A_{\mu} \rightarrow A_{\mu} - \xi_{\mu}$ under the secondary linearized 3D diffeomorphisms, and $\pi$ is similarly a scalar under the former and invariant under the later.
Since the Lagrangian (\ref{lagr2}) depends only directly on $\bar V_{\mu}$, it is manifestly invariant under a new additional $U(1)$ symmetry
\beq
A_\mu\rightarrow A_\mu+\partial_\mu \zeta, \quad \quad \pi\rightarrow \pi-\zeta,
\eeq
generalizing the result in the linearized theory Eq.~(\ref{gt}). Since the above Lagrangian is only linear in  $\dot {\bar V}_0$ it is in turn guaranteed to be only linear in $\ddot{\pi}$.  Any such interaction will lead to second order equations of motion for $\pi$.  However, one may worry about defining a conjugate momentum for $\pi$. That this can be done consistently,  
can be seen by noting that it is always possible to integrate by parts to put the action in a form where there are no more than single time derivatives acting on the fields in the Lagrangian.
To see this explicitly we note that the only term in the above Lagrangian where the $\ddot{\pi}$ terms arise, is in `gauged' Galileon interaction on the last line
\beq
\mathcal{L}_0=-\frac{1}{2}\sqrt{g}\, \nabla^\mu \bar V^\nu (\bar V_\mu \bar V_\nu+g\mn\bar V^\alpha \bar V_\alpha)\,,
\eeq
as well as in 
\ba
\mathcal{L}_1=-\sqrt{g}\, \tilde f^{\mu\nu}\(\nabla_\mu \bar V_\nu-g\mn \nabla^\alpha\bar V_\alpha \)=-\sqrt{g}\(\nabla_\nu\bar V_\mu\)(\tilde f^{\mu\nu}-\tilde f g^{\mu\nu})\,.
\ea
On integration by parts we see that the first term is equivalent to
\ba
\mathcal{L}_0=-\frac{1}{4}\sqrt{g} \, (\bar V^\alpha \bar V_\alpha) \nabla^\mu \bar V_\mu=-\frac 14  (\bar V^\alpha \bar V_\alpha) \partial_\mu(\sqrt{g}\, V^\mu)\,.
\ea
It is now easy to show that the would be problematic $\ddot{\pi}$ coming from the $\dot{\bar V}_0$ part is a total derivative and consequently can be removed by an integration by parts
\ba
\mathcal{L}_0 &=& - \frac 14  \Big[-\frac{1}{N^2}( \bar V_0-N^i\bar V_i)^2+\gamma^{ij}\bar V_i \bar V_i\Big] \Big[ \partial_0 \(\frac{\sqrt{\gamma}}{N} (-\bar V_0+N^i \bar V_i)\)+\partial_i ( \dots)\Big] \nn \\
&=& -  \frac{1}{12} \, \partial_0 \left( \sqrt{\gamma}\frac{(\bar V_0-N^i \bar V_i)^3}{N^3}\right) + \frac 14 \partial_0 \left(\frac{\sqrt{\gamma}}{N} ( \bar V_0-N^i \bar V_i)\gamma^{ij}\bar V_i \bar V_i \right) \nn \\
&+&\text{   terms with {\bf{no}} $\dot{\bar V}_0,\dot{N}$ or $\dot{N}^i$} ,
\ea
where $N$ is the ordinary lapse $g^{00}=-1/N^2$, $g_{0i}=N_i$ the associated shift and $g_{ij}=\gamma_{ij}$.
This is precisely for the same reason that the covariant DGP Galileon interaction $\Box \pi (\partial \pi)^2$ term gives well defined equations of motion. Indeed setting $A_{\mu}=0$ we see easily that this term is equivalent to the DGP interaction for $\pi$:
\ba
 (\bar V^\alpha \bar V_\alpha) \nabla^\mu \bar V_\mu|_{A_\mu=0}=\Box \pi (\p \pi)^2\,.
\ea
Implicitly these integrations by parts determine the analogue of the Gibbons-Hawking boundary terms for NMG.
A similar argument can be employed for $\mathcal{L}_1$, where all potentially problematic terms can be seen to be a total derivative:
\ba
\mathcal{L}_1&=&-\sqrt{\gamma}N_0 \(\frac{\dot N}{N}V_i N^i-V_0\dot N-V_i \dot N^i+\dot V_0+\cdots \)\frac{1}{N_0^2}\gamma^{ij}\tilde f_{ij}+\cdots\nn\\
&=&-\sqrt{\gamma}\gamma^{ij}\tilde f_{ij}\(\partial_0\(\frac{V_0}{N}\)-V_i\partial_0\(\frac{N^i}{N}\)\)+\cdots
\ea
so that this term is also free of any time derivative on the Lagrange multipliers,
\ba
\mathcal{L}_1=-\partial_0\(\frac{\sqrt{\gamma}}{N}\gamma^{ij}\tilde f_{ij}\(V_0-V_iN^i\)\)+\text{   terms with {\bf{no}} $\dot{\bar V}_0,\dot{N}$, $\dot{N}^i$,  $\dot{\tilde f}_{00}$ or $\dot{\tilde f}_{0i}$} .
\ea
Again the key feature to stress is that at the same time as removing the $\dot{V}_0$ terms we can also remove the $\dot{N}$, $\dot{N}^i$, $\dot{\tilde f}_{00}$ and $\dot{\tilde f}_{0i}$ terms ensuring that $N$, ${N}^i$, $\tilde f_{00}$ and $\tilde f_{0i}$ act as Lagrange multipliers for the normal 3D diffeomorphism Hamiltonian and momentum constraints. Similarly since there are by construction no $\dot{A}_0$ terms, then $A_0$ acts as the Lagrange multiplier for the $U(1)$ `Gauss law' constraint.

Putting this together, since all the double time derivative terms for $\pi$ can be removed by integration by parts, and by the same means the Lagrangian can be made independent of $\dot{N}, \dot{N^i},\dot{\tilde f}_{00}, \dot{\tilde f}_{0i},\dot{A}_0$, we are guaranteed that the equations of motion are well defined, the 7 constraints are preserved, and that there is a well-defined Hamiltonian formalism. The key point then is that following these steps the configuration space action can be written in a form in which it is a function of the 6+6+3+1=16 variables $g_{\mu\nu},f_{\mu\nu},A_{\mu},\pi$ each of which has well defined equations of motion. In addition there are 3+3+1=7 first class symmetries corresponding to 3D diffeomorphisms, secondary linearized 3D diffeomorphisms and the additional $U(1)$. Associated with each symmetry there is a gauge freedom and a constraint which allows us to remove 2 degrees of freedom per symmetry. Thus the total number of degrees of freedom non-perturbatively is 16 - 7 -7 =2. Since 2 is the correct number of physical polarizations of a massive spin 2 field in 3D this confirms the absence of the BD ghost to all orders, consistent with our previous arguments which were only valid in certain regimes.

\section{Explicit Computation of the Decoupling Limit at Nonlinear Order}

In this section we analyze the decoupling limit of NMG explicitly to cubic order. We show that  one indeed recovers a ghost-free theory, governed by the scale $\Lambda_{5/2}$. We do so by explicitly demonstrating how the terms corresponding to the $\Lambda_{9/2}$ and $\Lambda_{7/2}$ all vanish. We then generalize the results and construct a new class of theories which are ghost-free to cubic order.

\subsection{NMG at Nonlinear Order}

To obtain the NMG Lagrangian to cubic order one must integrate out the metric perturbation $h_{\mu\nu}$ from the equations of motion of NMG (\ref{eq1})-(\ref{eq2}) beyond the quadratic order in the Lagrangian. This is a rather lengthy computation and it is outlined in the appendix. The resulting cubic effective Lagrangian for $f_{\mu\nu}$ can be written in the following form,
\ba
\mathcal{L}^{(3)}_f &=& 2M_P(\sqrt{\gamma}R(\gamma))^{(3)}+\frac{1}{2}M_P f^{\mu\nu}\mathcal{E}_{\mu\nu}^{\rho\sigma} f_{\rho\sigma}
-\frac{M_P m^2}{4} ( f^2_{\mu\nu}-f^2 \nn \\ &+&\frac{5}{2}f^2_{\mu\nu}f- 2f^3_{\mu\nu}-\frac{1}{2}f^3).
\label{clag}
\ea
Here $\gamma_{\mu\nu}\equiv \eta_{\mu\nu}+f_{\mu\nu}$ is the analog of the full metric with $f_{\mu\nu}$ viewed as metric perturbation, and $M_P(\sqrt{\gamma}R(\gamma))^{(3)}$ denotes the cubic in $f_{\mu\nu}$ part of the corresponding Einstein-Hilbert action. At the quadratic level, we  of course recover the FP action; a remarkable peculiarity of the cubic part, however, is that it modifies GR not just by the graviton's potential, but also by the deformation of the nonlinear Einstein-Hilbert part itself.

According to the discussion of section 2, in a generic nonlinear extension of the FP action we should expect $\Lambda_{9/2}$ to be the scale, governing the dynamics of the helicity-0 mode. On the other hand, we have shown in the previous section that $\Lambda_{5/2}$ represents the lowest scale, by which the Galileon self-interactions of the scalar graviton are suppressed in NMG. The theory must therefore possess a special structure, which eliminates the scale $\Lambda_{9/2}$, as well as the intermediate scale $\Lambda_{7/2}\equiv(\sqrt{M_P}m^3)^{2/7}$, from its dynamics. To show this, let us plug the decomposition (\ref{nls}) (with the obvious change $h\to f$, while keeping the same notation for the vector and scalar modes) into the cubic Lagrangian Eq.~(\ref{clag}). In the limit where all scales greater than $\Lambda_{7/2}$ are sent to infinity, one recovers the following expression for the action,
\beq
\mathcal{L}^{dl}_{f} &=& -\frac{1}{2}\tilde{f}^{\mu\nu}\E^{\alpha\beta}_{\mu\nu} \tilde{f}_{\alpha\beta}-\frac{1}{4}F^{\mu\nu}F\mn+ 2\pi\Box\pi \nn \\ &+& \frac{1}{\Lambda_{9/2}^{9/2}}
\left(
2  \partial_\mu\partial_\nu\pi\partial^\mu\partial_\alpha\pi \partial^\nu\partial^\alpha\pi -
3  \Box\pi\partial_\mu\partial_\nu\pi\partial^\mu\partial^\nu\pi +
\left(\Box\pi\right)^3
\right)
\nn \\&+&
\frac{1}{\Lambda^{7/2}_{7/2}}(
7 \partial_\mu A_\nu \partial^\mu\partial^\alpha\pi \partial^\nu\partial_\alpha \pi -6 \partial_\mu A_\nu \partial^\mu\partial^\nu\pi \Box \pi
\nn \\ &-&4 \partial_\mu A^\mu \partial_\alpha\p_\beta\pi \partial^\alpha\p^\beta\pi +
3 \partial_\mu A^\mu \Box\pi \Box \pi
 ) \nn \\
&+&\(\frac{1}{\Lambda^{7/2}_{7/2}}\(\p^\mu A^\rho \p^\nu\p_\rho \pi+\p^\nu A^\rho \p^\mu\p_\rho \pi\)+\frac{1}{\Lambda^{9/2}_{9/2}}\p^\mu\p^\rho\pi\p^\nu\p_\rho\pi \)\E^{\alpha\beta}_{\mu\nu} \tilde{f}_{\alpha\beta}.
\label{lagr'''}
\eeq
Here $\tilde f =\hat f-2\pi\eta$ represents the familiar conformally shifted tensor mode, so that  in terms of the new fields the $U(1)$ gauge symmetry of (\ref{gt}) is given by,
\beq
\tilde f\mn\to \tilde f\mn+2 m\xi \eta\mn, \quad A_{\mu}\to A_{\mu}+\p_{\mu}\xi, \quad \pi\to\pi-m\xi.
\label{gi'}
\eeq
Under these transformations, the Lagrangian (\ref{lagr'''}) is invariant up to terms that vanish in the decoupling limit,
\beq
\delta_\xi \mathcal{L}^{dl}_{f} = \mathcal{O} \(\frac{\xi}{\Lambda^{5/2}_{5/2}}\).
\eeq

The $\pi$ self-interactions in the second line of Eq.~(\ref{lagr'''}) combine to a total derivative and therefore drop out. Furthermore, extracting a total derivative\footnote{The corresponding total derivative is given by \\ $ 7 \(  \p^\mu A^\nu\p_\mu\p^\alpha\pi\p_\nu\p_\alpha\pi -  \p^\mu A^\nu\p_\mu\p_\nu\pi\Box\pi -\frac{1}{2} \p^\mu A_\mu(\p_\alpha\p_\beta\pi)^2 +\frac{1}{2} \p^\mu A_\mu(\Box\pi)^2 \)/\Lambda^{5/2}_{5/2}$.} from the $\p A\p\pi\p\p\pi$-type operators and using the gauge invariance Eq.~(\ref{gi'}) to impose the condition $\p^\mu A_\mu=0$, (\ref{lagr'''}) can be rewritten as,
\beq
\mathcal{L}^{dl}_{f} &=& -\frac{1}{2}\tilde{f}^{\mu\nu}\E^{\alpha\beta}_{\mu\nu} \tilde{f}_{\alpha\beta}-\frac{1}{4}F^{\mu\nu}F\mn+ 2\pi\Box\pi +\frac{1}{\Lambda^{7/2}_{7/2}}\partial_\mu A_\nu \partial^\mu\partial^\nu\pi \Box \pi \nn \\ &+&\(\frac{1}{\Lambda^{7/2}_{7/2}}\(\p^\mu A^\rho \p^\nu\p_\rho \pi+\p^\nu A^\rho \p^\mu\p_\rho \pi\)+\frac{1}{\Lambda^{9/2}_{9/2}}\p^\mu\p^\rho\pi\p^\nu\p_\rho\pi \)\E^{\alpha\beta}_{\mu\nu} \tilde{f}_{\alpha\beta}.
\eeq
The remaining interactions can be removed by a field redefinition of the helicity-0 and helicity-2 fields,
\ba
\pi\to\pi-\frac{1}{4\Lambda^{7/2}_{7/2}}\partial_\mu A_\nu \partial^\mu\partial^\nu\pi; ~  \tilde{f}_{\mu\nu} \to \tilde{f}_{\mu\nu}&+&\frac{1}{\Lambda^{7/2}_{7/2}}\(\p_\mu A^\rho \p_\nu\p_\rho \pi+\p_\nu A^\rho \p_\mu\p_\rho \pi\)\nn \\ &+&\frac{1}{\Lambda^{9/2}_{9/2}}\p_\mu\p^\rho\pi\p_\nu\p_\rho\pi,
\ea

As promised, all the dangerous interactions associated with the $\Lambda_{9/2}$ and $\Lambda_{7/2}$ scales have dropped-out in this limit. This leaves $\Lambda_{5/2}$ to control the nonlinear dynamics of NMG, as shown in subsection 3.1.

It is important to note the significance of the $U(1)$ gauge invariance in the preceding discussion. In fact, precisely due to this freedom, any operator of the type $\p A\p^2\pi\p^2\pi$ at the scale $\Lambda_{7/2}$ can be removed (up to a total derivative) by a redefinition of the helicity-0 graviton. This means that once the problematic $(\p^2\pi)^3$-type operators at the scale $\Lambda_{9/2}$ are removed by a judicious choice of the coefficients of the different terms in the Lagrangian, the $\Lambda_{7/2}$ scale becomes automatically removable, leading to the decoupling limit, governed by $\Lambda_{5/2}$\footnote{Mathematically, this traces back to the fact that the $U(1)$ transformation of the scalar in (\ref{gt}) includes a factor of the mass $m$, connecting the scale $\Lambda_{9/2}$ to $\Lambda_{7/2}=m/\Lambda_{9/2}$. Because of this, tuning the theory so as to remove the former scale automatically results in the removal of the latter one, as was for instance encountered in \cite{drg2,drgt}.}.

\subsection{General Cubic-Order Ghost-Free Massive Gravity}

Finally,  we extend NMG at cubic order considering a more general set of derivative self-interactions of the graviton. The generic type of cubic Lagrangians we wish to consider is written in terms of the Einstein-Hilbert term (with possibly a non-standard coefficient) together with reparametrisation invariance-violating contributions involving the metric perturbation $h\mn=g\mn-\eta\mn$,
\begin{eqnarray}
\label{eqn:generalLag}
\mathcal{L} &=& M_{P}~\left[\rho (\sqrt{g} R)^{(3)} - \frac{m^2}{4}\left(h_{\mu\nu}^2 - h^2 + c_1 h_{\mu\nu}^3 + c_2 h h_{\mu\nu}^2 + c_3 h^3 + \ldots \right) \right. \\\nonumber
&~~~& \left. + \alpha h^{\mu\nu}G\mn^{(1)} + \beta_1 h^{\mu\rho}h_\rho^\nu G\mn^{(1)} +\beta_2 h h^{\mu\nu}G^{(1)}\mn+ \beta_3 h\mn h^{\mu\nu}G + \beta_4 h^2G^{(1)} +\ldots \right]
\end{eqnarray}
where the ellipses denote possible higher order terms \footnote{All terms except the first one in the second line of (\ref{eqn:generalLag}) can in fact be removed at the cubic order by a \emph{nonlinear} redefinition of the metric perturbation $h_{\mn}$; in particular, setting $h_{\mn} \to h_{\mn}+ \beta_1 h_\mu^\rho h_{\rho\nu}+\beta_2 h h_{\mu\nu}+ \beta_3 h^{\alpha\beta} h_{\alpha\beta}\eta_{\mn}+ \beta_4 h^2\eta_{\mn}$, will eliminate the $hhG^{(1)}$-type terms, while generating additional $\beta$-dependent polynomial contributions. This leads to an alternative, equivalent formulation of the cubic theory; we will however choose to work with the original form of the action (\ref{eqn:generalLag}) below.}.
In order to investigate the relevant degrees of freedom of this theory, we employ again the \stu decomposition, Eq.~(\ref{nls}).
We begin by investigating the $\Lambda_{9/2}$ - limit,
\beq
\label{eqn:Lambda92Limit}
m \to 0, ~~~\mpl \to \infty, ~~~  {\rm keeping\ } \Lambda_{9/2} \equiv  (m^4 \sqrt{\mpl} )^{2/9}
 ~~{\rm fixed}\,.
\label{declim3}
\eeq
Let us for definiteness concentrate on the particular value $\rho=2$ for the coefficient in front of the Einstein-Hilbert term, since it corresponds to the case of NMG as shown above. Then, in order to recover the FP Lagrangian at the linearized level, $\alpha=1/2$ should hold. Expanding the Lagrangian of Eq.~(\ref{eqn:generalLag}) up to cubic order and performing the usual rescaling $\hat{h}\mn\rightarrow \tilde{h}\mn + 2\eta\mn \pi$, we obtain,
\beq
\mathcal{L}_{9/2} &=& -\frac{1}{2}\tilde {h}^{\mu\nu}\E^{\alpha\beta}_{\mu\nu} \tilde{h}_{\alpha\beta} \nn \\&+& \frac{3}{2}\pi\Box\pi + \frac{1}{\Lambda_{9/2}^{9/2}}
\left(
\alpha_1  \partial_\mu\partial_\nu\pi\partial^\mu\partial_\alpha\pi \partial^\nu\partial^\alpha\pi +
\alpha_2  \Box\pi\partial_\mu\partial_\nu\pi\partial^\mu\partial^\nu\pi +
\alpha_3  \left(\Box\pi\right)^3
\right), \nn
\eeq
where the corresponding coefficients are given as follows,
\beq
\label{eqn:pipipiCoefficients}
\alpha_1 &=& -2-2c_1 -4\beta_1,\\\nonumber
\alpha_2 &=& 2 -2 c_2 + 4 \beta_1 - 4 \beta_2 + 8 \beta_3,\\\nonumber
\alpha_3 &=& -2 c_3 + 4 \beta_2 + 8 \beta_4.
\eeq
In order to avoid the propagation of ghosts as discussed above,  we choose these coefficients such that the three terms form a total derivative (or vanish) and therefore have no effect on the equations of motion. The relevant conditions therefore are,
\beq
\alpha_1=-\frac{2}{3}\alpha_2=2\alpha_3.
\eeq
NMG represents one example of a theory for which these conditions hold.

\vspace{10pt}

\noindent {\bf Acknowledgements:}
We would like to thank Gaston Giribet for useful correspondence. CdR is funded by the SNF, the work of GG was supported by NSF grant PHY-0758032, DP is supported by the Mark Leslie
Graduate Assistantship at NYU and IY is supported by the James Arthur fellowship. CdR and AJT would like to thank NYU for hospitality whilst part of this work was being completed.

\renewcommand{\theequation}{A-\Roman{equation}}
\setcounter{equation}{0}
\section*{Appendix A}

In this appendix we briefly outline the procedure of integrating out the metric perturbation $h_{\mu\nu}$ in (\ref{eq1})-(\ref{eq2}) in order to obtain the effective equations of motion, quadratic in $f_{\mu\nu}$ (or equivalently, the cubic effective Lagrangian for $f\mn$).

For this, we can use the expression for $\sqrt{g}G^{\mu\nu}$ found from (\ref{eq1}) in the equation for the auxiliary field (\ref{eq2}), and take into account the first-order relation $h_{\mu\nu}=f_{\mu\nu}$ to rewrite everything in terms of $f_{\mu\nu}$ only. Specifically, the latter relation is used for rewriting the (linearized) Einstein and Ricci tensors in terms of $f_{\mu\nu}$ in the first line of (\ref{eq1}), as well as the (linearized) quantities such as  $\sqrt{g}$, $g^{\mu\nu}$ and the covariant derivatives in the second line of (\ref{eq1}) and in the second term of Eq. (\ref{eq2})\footnote{Note that we do not have to go beyond the first order in expressing $h_{\mu\nu}$ in terms of $f_{\mu\nu}$ for obtaining the second-order equations for the latter field.}. After a rather involved but straightforward computation, one arrives at the following effective equation for $f_{\mu\nu}$
\beq
(\sqrt{\gamma}G^{\mu\nu}(\gamma))^{(1)}+2 (\sqrt{\gamma}G^{\mu\nu}(\gamma))^{(2)}+\frac{1}{2}m^2(f^{\mu\nu}-\eta^{\mu\nu}f)~~~~\nonumber \\ +m^2 \left( \frac{5}{8} \eta^{\mu\nu}f^{\alpha\beta}f_{\alpha\beta}-\frac{3}{8}\eta^{\mu\nu}f^2-\frac{3}{2}f^{\mu\alpha}f_\alpha^\nu+\frac{5}{4}f^{\mu\nu}f \right)=0,
\label{qe}
\eeq
where $\gamma_{\mu\nu}\equiv\eta_{\mu\nu}+f_{\mu\nu}$ and $(\sqrt{\gamma}G^{\mu\nu}(\gamma))^{(n)}$ denotes the contribution to the corresponding quantity of order $n$ in $f_{\mu\nu}$ \footnote{The perturbative equation of motion for a massless GR graviton $h_{\mu\nu}$ is of the form
$(\sqrt{g}G^{\mu\nu}(g))^{(1)}+(\sqrt{g}G^{\mu\nu}(g))^{(2)}+ ... = 0$. It is crucial that the indices are raised in the latter equation; since $h_{\mu\nu}$ is defined as the perturbation of the metric $g_{\mu\nu}$ (and not its inverse), the variation of the Einstein-Hilbert action w.r.t $h_{\mu\nu}$ yields the Einstein's equations with \textit{upper} indices, expressed through the metric perturbation.}. More explicitly, the first two terms of (\ref{qe}) involve the following expressions,
\beq
(\sqrt{\gamma}G^{\mu\nu}(\gamma))^{(1)}=\mathcal{E}^{\mu\nu}_{\rho\sigma} f^{\rho\sigma},\nonumber
\eeq
\ba
(\sqrt{\gamma}G^{\mu\nu}(\gamma))^{(2)}=\frac{1}{4}(\partial^\rho f \partial^\mu f^\nu_\rho+\partial^\rho f \partial^\nu f^\mu_\rho)-\frac{1}{4}\partial^\rho f \partial_\rho f^{\mu\nu}-\frac{1}{2}(\partial^\mu f^{\nu\rho} \partial^\sigma f_{\sigma\rho}+\partial^\nu f^{\mu\rho} \partial^\sigma f_{\sigma\rho})
\nn \\+ \frac{1}{2}\partial^\rho f _\rho^\sigma\partial_\sigma f^{\mu\nu}-\frac{1}{2}\partial^\rho f^{\mu\sigma} \partial_\sigma f^\nu_\rho+\frac{1}{2}\partial^\rho f^{\mu\sigma} \partial_\rho f^\nu_\sigma +\frac{1}{4}\partial^\mu f^{\rho\sigma} \partial^\nu f_{\rho\sigma}-\frac{1}{2}f^{\rho\sigma}(\partial_\rho\partial^\mu f^\nu_\sigma +\partial_\rho\partial^\nu f^\mu_\sigma) \nn \\
+\frac{1}{2}f^{\rho\sigma}\partial_\rho\partial_\sigma f^{\mu\nu}+\frac{1}{2}f^{\rho\sigma}\partial^\mu\partial^\nu f_{\rho\sigma}
-\frac{1}{2}\eta^{\mu\nu}[\partial^\rho f \partial^\sigma f_{\rho\sigma}-\frac{1}{4}\partial^\rho f\partial_\rho f -\partial^\rho f_\rho^\sigma\partial^\alpha f_{\alpha\sigma}-\frac{1}{2}\partial^\rho f^{\sigma\alpha}\p_\sigma f_{\rho\alpha}\nn \\+\frac{3}{4}\p^\alpha f^{\rho\sigma} \p_\alpha f_{\rho\sigma}
-2 f^{\rho\sigma}\partial_\rho\p^\alpha f_{\sigma\alpha}+f^{\rho\sigma}\p_\rho\p_\sigma f+f^{\rho\sigma}\Box f_{\rho\sigma}]+\frac{1}{2} f^{\mu\nu}\p^\rho\p^\sigma f_{\rho\sigma}-\frac{1}{2}f^{\mu\nu}\Box f+ \nn \\ \frac{1}{4} f (\p^\mu\p^\rho f_\rho^\nu  +\p^\nu\p^\rho f_\rho^\mu -\p^\mu\p^\nu f -\Box f^{\mu\nu}+\eta^{\mu\nu}\Box f-\eta^{\mu\nu}\p^\rho \p^\sigma f_{\rho\sigma})-\frac{1}{2} f^{\mu\rho}(\p_\rho\p^\sigma f_\sigma^\nu  +\p^\nu\p^\sigma f_{\sigma\rho} \nn \\ -\Box f^\nu_\rho-\p^\nu\p_\rho f)-\frac{1}{2} f^{\nu\rho}(\p_\rho\p^\sigma f_\sigma^\mu+\p^\mu\p^\sigma f_{\sigma\rho}-\Box f^\mu_\rho-\p^\mu\p_\rho f),\nonumber
\ea
where all indices are contracted with the flat metric.

It is now straightforward to obtain the cubic Lagrangian which yields (\ref{qe}) as the equation of motion for $f\mn$,
\beq
\mathcal{L}^{(3)}_f&=& 2M_P(\sqrt{\gamma}R(\gamma))^{(3)}-M_P(\sqrt{\gamma}R(\gamma))^{(2)}
-\frac{M_Pm^2}{4} ( f^2_{\mu\nu}-f^2 \nn \\ &+& \frac{5}{2}f^2_{\mu\nu}f- 2 f^3_{\mu\nu}-\frac{1}{2}f^3 ).
\label{applagr}
\eeq
Here, the notation $f^2_{\mu\nu}\equiv f^{\mu\nu}f_{\mu\nu}$ and $f^3_{\mu\nu}\equiv f^{\mu\alpha}f_{\alpha\beta}f^\beta_\mu$ has been used.

\end{document}